\documentclass{ws-procs975x65}
\usepackage{amsfonts}
\usepackage{amsmath}
\usepackage{amsbsy}
\usepackage{amssymb}

\font\grb=eurb10

\def\bphi{\hbox{\grb\char'047}\,}

\def\vpint{\mathop{\scriptstyle{\mathbf{\diagup}}\hskip-2ex \displaystyle{\int}}}

\begin{document}

\title{SUPERPOSITION OF FIELDS OF TWO REISSNER - NORDSTR\"OM SOURCES}

\author{G.A.~Alekseev}

\address{Steklov Mathematical
Institute, Gubkina 8, Moscow 119991, Moscow, Russia\\
\email{G.A.Alekseev@mi.ras.ru}}

\author{V.A.~Belinski}
\address{INFN, Rome University "La Sapienza",\,\,\, 00185 Rome, Italy,\\
ICRANet, Piazzale della Repubblica, 10, 65122 Pescara, Italy\\
and IHES, F-91440 Bures-sur-Yvette, France\\
\email{belinski@icra.it}}

\begin{abstract}
In this paper we present a 5-parametric family of static asymptotically flat solutions for the superposed gravitational and electromagnetic fields of two Reissner-Nordstr\"om sources with arbitrary parameters --- masses, charges and separating distance. A procedure for solving of the linear singular integral equation form of the electrovacuum Einstein - Maxwell equations for stationary axisymmetric fields is described in detail. The 4-parametric family of equilibrium configurations of two Reissner-Nordstr\"om sources (one of which should be a black hole and another one  --  a naked singularity) presented in our recent paper \cite{Alekseev-Belinski:2007} arises after a restriction of the parameters of the 5-parametric solution presented here by the equilibrium condition which provides the absence in the solution of conical points on the symmetry axis between the sources.

\end{abstract}


\section*{Introduction}
In our recent short paper \cite{Alekseev-Belinski:2007} we presented (in a surprisingly simple form) an exact 4-parametric family of static asymptotically flat solutions of electrovacuum Einstein - Maxwell equations which describes the equilibrium configurations of two (nonrotating) charged massess in General Relativity. In this paper we present a more general, 5-parametric solution of these equations which represent a nonlinear superposition of fields of two Reissner-Nordstr\"om sources with arbitrary mass and charge parameters and arbitrarily chosen separating distance and describe the procedure for a construction of a superposition of these fields. In the subsequent sections, we use a special divergent form of the reduced Einstein - Maxwell equations to derive the Komar-like integrals for the total gravitational mass and charge of this field and calculate  the physical masses and charges of the sources defined as the additive inputs of each source in the total gravitational mass and total charge of the system. The expression of our solution in terms of these physical parameters simplifies it considerably. Then we  determine the constraint which should be imposed on the parameters of this superposition of fields  to provide the absence in the remaining 4-parametric solution of any non-physical singularities such as the conical points on the axis outside the sources.
{\it This constraint plays the role of the condition for equilibrium  of these  sources in their common gravitational and electromagnetic fields.}

The 5-parametric solution, presented in this paper, was constructed using the monodromy transform approach \cite{Alekseev:1985,Alekseev:1988} which give rise to a reformulation of the Einstein - Maxwell equations for stationary axisymmetric fields in terms of equivalent system of linear singular integral equations. As it is explained below,  the  functional parameters (monodromy data) in the kernels of these integral equations for our solution are chosen as analytically matched, rational functions of the spectral parameter. This choice of the monodromy data means that the solution possess \textit{rational axis data}, i.e. that the values of all components of metric and electromagnetic potential, expressed in terms of cylindrical Weyl coordinates $(\rho,z)$, on the axis of simmetry $\rho=0$ are rational functions of $z$. This means also that our solution virtually is the soliton solution on the Minkowski background in the sense, that it consists of a pure soliton part and its analytical continuation in the space of parameters.\footnote{This analytical continuation is similar to that which connects different parts of the Kerr-Newman family of solutions corresponding to the field of a naked singularity and the black hole solution.}

\subsection*{On the integral equation methods}
Different approaches to solution of integrable space-time symmetry reductions of the Einstein's field equations, which began to develop
about thirty years ago, gave rise to different formulations (generally non-equivalent to each other) of the integral equation methods, which allow to calculate the solutions of these field equations solving some systems of linear singular integral equations.

The first reformulation of symmetry reduced Einstein equations in terms of a system of linear singular integral equations for vacuum gravitational fields (together with a construction of vacuum solitons) was proposed in the framework of the {\it inverse scattering approach} \cite{Belinski-Zakharov:1978}{}. This construction was based on a formulation of equivalent Riemann - Hilbert problem for the $2\times 2$-matrix functions on the spectral plane which gave rise to some system of linear singular integral equations. It is important, that the solutions of this equations possess the character of {\it solution generating transformations}, because the matrix kernel of these integral equations includes, as functional parameters, the components of an arbitrarily chosen vacuum metric which serves as the background for   solitons, while the generating solution itself can be considered as describing some nonlinear perturbation of this background.

In the framework of another, {\it group-theoretic approach}, for construction of the solution generating transformations corresponding to the elements of the infinite dimensional algebra of internal symmetries of electrovacuum Einstein-Maxwell equations for stationary axisymmetric  fields (found by Kinnersley and Chitre\, \cite{Kinnersley-Chitre:1977}{}), Hauser and Ernst \cite{Hauser-Ernst:1979} reduced these equations  to a homogeneous Hilbert problem for $3\times 3$-matrix functions of an auxiliary complex parameter and then, to the corresponding $3\times 3$-matrix linear singular integral equations. This construction assumed an additional constraint imposed on the class of solutions, which means that only those solutions are considered which possess (locally) a regular behaviour of fields near the axis of symmetry. The integral equations \cite{Hauser-Ernst:1979} have rather complicate matrix kernel which construction includes a calculation of matrix exponents of the elements of Kinnersley and Chitre algebra, represented by some algebraically defined
holomorphic $3\times 3$-matrix functions of the mentioned above auxiliary complex parameter. Later, fixing the choice of the seed solution by the simplest one, Sibgatullin \cite{Sibgatullin:1984} reduced the  matrix integral equations \cite{Hauser-Ernst:1979} of Hauser and Ernst to a  simpler scalar linear singular integral equation with some "normalization" condition imposed additionally on its solutions. The kernel of this scalar integral equation was expressed explicitly in terms of the values of the Ernst potentials on the axis of symmetry. This integral equation was actively used during the last one and a half decades by Sibgatullin and others for mostly formal calculation of asymptotically flat solutions for various particular choices of (asymptotically flat) rational axis data for the Ernst potentials.

The {\it monodromy transform approach}\cite{Alekseev:1985,Alekseev:1988}  does not follow the ideology of the matrix Riemann-Hilbert problems, however, it is also based on some ideas of the modern theory of integrable systems, analytical theory of differential equations and the theory of linear singular integral equations. For physically different classes of vacuum and electrovacuum fields with two commuting isometries (stationary axisymmetric fields, plane and cylindrical waves, inhomogeneous cosmological solutions and some others) this approach suggests rather simple general construction of the coordinate-independent functional parameters (called as monodromy data) which characterize uniquely every local solution. The problem of constructing solutions for given monodromy data (the inverse problem of the monodromy transform) gave rise to a system of linear singular integral equations which differs essentially from the integral equations mentioned above. A specific structures of the (scalar) kernels of these integral equations and of the integration path on the spectral plane allow to describe all degrees of freedom of the gravitational and electromagnetic fields and make these integral equations {\it equivalent} to the symmetry reduced Einstein - Maxwell equations. For stationary axisymmetric fields satisfying the regularity axis condition, the integral equations \cite{Alekseev:1985,Alekseev:1988} simplify considerably. A general scheme for constructing of solutions with any  {\it rational analytically matched monodromy data} (or, equivalently, of solutions for any rational, not only asymptotically flat, axis data) was described earlier in detail (see \cite{Alekseev:1988, Alekseev-Garcia:1996} and the references therein).

\subsection*{Solitons on the Minkowski background and  rational axis data.}
A discovery of existence of pure gravitational solitons and of the ways for their generating on arbitrarily chosen vacuum backgrounds -- the dressing methods\cite{Belinski-Zakharov:1978}(see also\, \cite{Belinski-Verdaguer:2001}), together with initiated by these results developments of the similar methods for Einstein - Maxwell fields\cite{Alekseev:1980}{}, gave us a powerful tool for construction of a large variety of solutions of Einstein's field equations and for nonlinear superposition of fields of certain kinds of sources with various external fields.

On the other hand, later developments of the integral equation methods briefly described above suggested a simple idea to calculate the solutions of Einstein and Einstein - Maxwell equations using these integral equations with the simplest choice of functional parameters (the contour data, or the axis data for the Ernst potentials, or analytically matched monodromy data) in their kernels as rational functions of their arguments. This also leads to a construction of large families of exact solutions with arbitrary (finitely large) number of free parameters, but without any freedom in the choice of the background solution. This gives rise to the obvious questions concerning a comparison of these two constructions.

First of all, it is clear, that the asymptotically flat rational-data solutions can be only a very special case of soliton solutions corresponding to a particular choice of the background for solitons. This can be the Minkowski space-time or any other soliton solution on this background. Indeed, a direct calculation shows, that vacuum solitons\, \cite{Belinski-Zakharov:1978} and electrovacuum solitons \cite{Alekseev:1980}{}, both generated on the Minkowski background, represent the asymptotically flat solutions with rational axis data. These solitons have as many of free parameters per one simple pole of the dressing matrix, as it is necessary to describe an arbitrary chosen rational asymptotically flat axis data with a twice lower number (for vacuum case) or with the same number of poles in this data.\footnote{It is necessary to note, however, that the poles of the dressing matrices for solitons coincide with the poles of the solutions of the integral equations, but they do not coincide in general, with the poles of the axis data for the Ernst potentials.} Of course, the coincidence of the number of parameters itself does not mean that these classes of solutions coincide. Moreover, some doubts in this coincidence can arise from that fact that  the construction \cite{Alekseev:1980}{} of electrovacuum solitons (in contrast to vacuum soliton generating technique \cite{Belinski-Zakharov:1978}{}) allows to generate the solitons with arbitrarily located complex poles of the dressing matrix and with the  complex conjugated poles of the inverse matrix, while the solitons with real poles do not arise in this technique. Of course, one can try to obtain the electrovacuum solitons with real poles as limiting cases of solitons with complex poles, however, this leads to solutions with coinciding real poles of dressing matrix and its inverse, which have less number of free parameters per each pole than the solitons with complex poles.

To clarify this situation, it is worth to consider a one-soliton solution on the Minkowski background, which coincides with the over-extreme part of the well known Kerr-Newman solution corresponding to a naked singularity. The under-extreme part of this solution, which has a horizon and corresponds to a black hole, can be described in terms of the soliton generating technique by a dressing matrix with one real pole. However, as it is well known, the under-extreme part of the Kerr-Newman family can be obtained by a simple analytical continuation from its over-extreme part in the space of its parameters. It is very likely, that the similar analytical continuation is possible not only for a one-soliton solution on the Minkowski background, but for any number of solitons as well. Indeed, solving the integral equations for the (static) two-pole monodromy data, we have found that the part of our static 5-parametric solution, which corresponds to a pair of naked Reissner-Nordstr\"om singularities, coincides with the static subfamily of two-soliton solutions on the Minkowski background, while the remaining part of this family is connected with this two-soliton solution by the similar analytical continuations in the space of parameters as one uses in the one-soliton case to connect different parts of the Kerr-Newman family of solutions. The motivation given above shows that there is the most reason to expect that the class of solutions of the integral equations for asymptotically flat rational axis data consists of the already known solitons on the Minkowski background together with their analytical continuations in the space of parameters (which should be supplied also, for completeness, by certain limiting solutions corresponding to multiple poles in the dressing matrices). On the one hand, in view of the above considerations, this means that there is no any reason to consider every formally calculated solution of the integral equations for rational asymptotically flat axis data as some new one describing the "extended" or "generalized" solitons, because it would be curiously enough to call similarly the under-extreme part of the Kerr-Newman family of solutions which is a simple analytically continuation in the parameter space of its over-extreme part.

On the other hand, the above motivation does not mean at all, that there is no sense to use the integral equation methods for calculation of these soliton solutions starting directly from some asymptotically flat rational axis data. It is worth to mention here, that the application of the soliton generating technique leads to some specific parameterization of the constructing solutions which differs significantly from that which arise for the solutions constructed from the corresponding solutions of the integral equations. Each of these parameterizations can occur to be more or less useful in different considerations of this class of soliton (or rational-data) solutions. For example, it seems, that just the existence of the analytical continuation in the parameter space discussed above can be proved for N-soliton solution more easily if we use its expression which arises from the solution of the integral equations, but considering the general twelve-parametric two-soliton solution, we find that it has the most compact form just in the soliton parameterization\cite{Alekseev:1988}.

This concludes our very brief reminding of some fragments of the history of the methods and of some interrelations between the solitons and solutions with rational axis data derived from the integral equation methods. At the end of this Introduction, we sketch out the monodromy transform approach and the integral equation method which we use  for construction of  our 5-parametric solution for a superposition of fields of two Reissner-Nordstr\"om sources. In all necessary details, a theory of this method and a general scheme for construction of solutions with rational, analytically matched monodromy data, have been developed  long ago and described in the papers cited above with some additional useful references therein.

\subsection*{Monodromy transform approach}

This approach is based a) on the parametrization of the space of local solutions of the symmetry reduced Einstein - Maxwell equations by the  monodromy data -- a set of coordinate independent functions of a spectral parameter, which determine the branching properties on the spectral plane of the fundamental solution of associated linear system, and b) on the reformulation of these equations in terms of a system of linear singular integral equations. For stationary axisymmetric fields, this approach
allows to construct  nonlinear superpositions of electrovacuum fields of different sources characterized by analytically matched, rational monodromy data \cite{Alekseev:1988} (see also the Appendix in \cite{Alekseev-Garcia:1996}). In the following sections, we demonstrate at first that the external field of a single Reissner-Nordstr\"om source is characterized just by this kind of the monodromy data functions which have on the spectral plane one simple pole and vanish at infinity. For superposition of fields of two such sources we choose the rational monodromy data functions as the sums of two poles with such coefficients which guarantee that the solution is static. Given this monodromy data, we describe step by step the construction of the corresponding solution.

\subsection*{The space of local solutions and its  parameterization by monodromy data}
For electrovacuum Einstein - Maxwell fields depending only on two space-time coordinates, in the entire space of local solutions, which are analytical near some initial point and take at this point (the point of "normalization") some "standard" values, every local solution with the metric $g_{ik}$ and electromagnetic potential $A_i$ can be  characterized uniquely by the monodromy data which consist of four coordinate-independent holomorphic functions of the spectral parameter $w$:
\begin{equation}\label{MData}
\{g_{ik}(x^1,x^2),A_i(x^1,x^2)\}\quad\longleftrightarrow
\quad\{\mathbf{u}_\pm (w),\mathbf{v}_\pm (w)\}
\end{equation}
This monodromy data are defined as a complete set of independent functions which characterize the branching properties of the corresponding fundamental solution of the associated linear system on the spectral plane at its four singular points. This data are defined uniquely for any local solution  and for arbitrary choice of these functions there always exists a unique local solution of electrovacuum Einstein - Maxwell equations with given monodromy data.

This construction is useful as far as some effective methods can be developed for explicit constructing of local solutions for given monodromy data. Then, given an explicit local solution, the corresponding global solution  can be determined using its analytical continuation to other space-time regions where this solution may reveal also some regular behaviour or approaches various types of singularities.

\subsection*{The master system of linear singular integral equations}
The key point of the use of the monodromy data is the existence of some system of linear singular integral equations whose kernels and right hand sides are expressed algebraically in terms of the monodromy data  and whose solution determines (by means of some quadratures) all components of metric and electromagnetic potential.

In general, the structure of this system of linear singular integral equations is rather complicate. The singular integrals are defined on the contour $L$ on the spectral plane  which consists of two disconnected parts $L=L_++L_-$. The locations of their endpoints depend on the space-time coordinates and coordinates of the initial point which enter also the integrands as parameters. In particular, $L_+$ goes from $w=\xi_o$ to $w=\xi$ and $L_-$ goes from $w=\eta_o$ to $w=\eta$, where, for example,  for stationary axisymmetric fields in terms of Weyl coordinates $\xi=z+i\rho$, $\eta=z-i\rho$, but for plane waves $\xi=x+t$, $\eta=x-t$) and $(\xi_o,\eta_o)$ correspond to the initial point.

The monodromy data as well as the unknown functions in these integral equations are defined in two disconnected regions of the spectral plane -- the neighbourhoods of $L_+$ and $L_-$, where they are represented by pairs of functions
($\mathbf{u}_+$, $\mathbf{u}_-$), ($\mathbf{v}_+$, $\mathbf{v}_-$) and ($\bphi_+$, $\bphi_-$) respectively. It is clear that these equations can not be solved for arbitrarily chosen monodromy data. However, for some classes of fields, such as, for example, stationary axisymmetric fields with a regular axis of symmetry  considered in this paper, these integral equations can be simplified considerably and admit infinite hierarchies of multiparametric families of explicit solutions.

\subsection*{Monodromy data for stationary fields with a regular symmetry axis}
For stationary axisymmetric fields, it is typical that physical and geometrical formulations of various problems (as, for example, in the case of asymptotically flat fields) imply that at least some part of the axis of symmetry is free of the field sources and therefore, a behaviour of metric and matter fields in the neighbourhood of this part of the axis should be regular. For such fields, if we choose the initial point (the point of normalization) of a solution on such regular part of the axis,  the initial points of the contours $L_+$ and $L_-$ coincide and instead of two disconnected contours we obtain one simple curve $L=L_++L_-$. Changing preliminary the direction of integration on $L_-$, we obtain one contour  where the system of singular integral equations is defined. It starts at the point $w=\eta$, goes through the initial point $w=z_0$ and ends at $w=\xi$. On this contour the monodromy data and the unknown variable in the integral equations are represented by twice lower number of holomorphic functions because for these fields we have
\begin{equation}\label{MatchedData}
\mathbf{u}_+(w)=\mathbf{u}_-(w)\equiv \mathbf{u}(w),\quad
\mathbf{v}_+(w)=\mathbf{v}_-(w)\equiv \mathbf{v}(w),\quad
\bphi_+(w)=\bphi_-(w)\equiv \bphi(w)
\end{equation}
Usually, we call these conditions as the regularity axis condition, however we note that these conditions guarantee only a regular local behaviour of fields near the axis, but they do not exclude the presence of some non-curvature singularities, such as conical points  on the axis or closed time-like curves near it. Thus, for stationary axisymmetric electrovacuum fields with a regular axis of symmetry the monodromy data are represented only by two arbitrary holomorphic functions $\mathbf{u}(w)$ and $\mathbf{v}(w)$. We recall also that $\mathbf{v}(w)$ is "responsible" for a presence of electromagnetic field, so that for vacuum $\mathbf{v}(w)\equiv 0$ and the space of solutions of vacuum stationary axisymmetric fields near the regular part of the axis of symmetry is parameterized by the  monodromy data consisting of one  holomorphic function $\mathbf{u}(w)$.

Another important property of the stationary axisymmetric electrovacuum fields with the regular axis of symmetry is that
any solution can be characterized by the finite values of its metric components and potentials (e.g., of the complex Ernst potentials $\mathcal{E}(\rho,z)$, $\Phi(\rho,z)$) on the regular part of the axis of symmetry. In this case, the monodromy data (\ref{MatchedData}) can be related to the values $\mathcal{E}(z)$, $\Phi(z)$ on the axis:
\begin{equation}\label{AxisData}
\mathcal{E}(\rho=0,z)=\mathcal{E}_0-2 i(z-z_0)\mathbf{u}(w=z),\quad \Phi(\rho=0,z)=\Phi_0+2 i(z-z_0)\mathbf{v}(w=z),
\end{equation}
where $z$ is the Weyl coordinate along the axis, $z_0$ is a coordinate of the initial point on this axis, $\mathcal{E}_0$ and $\Phi_0$ are the "normalized" values of the Ernst potentials at the initial point, for which we usually put $\mathcal{E}_0=1$ and $\Phi_0=0$.

\subsection*{The conditions for asymptotically flat and static  fields}
The expressions (\ref{AxisData}) allow us to relate the structure of the monodromy data (\ref{MatchedData}) with some physical and geometrical properties of fields. In particular, for any asymptotically flat field  $\mathbf{u}(w)$ and $\mathbf{v}(w)$ should be holomorphic at $w=\infty$ and
\[\mathbf{u}(w)\to 0\qquad\hbox{and}\qquad \mathbf{v}(w)\to 0\qquad\hbox{for}\qquad
w\to\infty.
\]
Moreover, in this case, the coefficients of expansions of these functions in the inverse powers of $w$ for $w\to\infty$ can be simply related to the multipole moments of this asymptotically flat field. Therefore, the multipole structure of the field can be determined in advance by the appropriate choice of the monodromy data.

For static fields, $\mathcal{E}$ should be real, while   $\Phi$ should be real for pure electric fields and imaginary for pure magnetic fields and therefore, $\mathbf{u}(w)$ and  $\mathbf{v}(w)$ should satisfy
\[\mathbf{u}^\dagger(w)=-\mathbf{u}(w)\qquad\hbox{and}\qquad
\mathbf{v}^\dagger(w)=\mp\mathbf{v}(w)
\]
where $\mathbf{u}^\dagger(w)\equiv \overline{\mathbf{u}(\overline{w})}$ and $\mathbf{v}^\dagger(w)\equiv \overline{\mathbf{v}(\overline{w})}$ and a bar means a complex conjugation.

\subsection*{Exact solutions with rational monodromy data}

An infinite hierarchies of solutions of the Einstein - Maxwell equations  can be calculated explicitly if we choose the analytically matched monodromy data (\ref{MatchedData}) to be rational functions of the spectral parameter:
\[\mathbf{u}(w)=\dfrac{U(w)}{Q(w)},\qquad
\mathbf{v}(w)=\dfrac{V(w)}{Q(w)}
\]
where the functions $U(w)$, $V(w)$ and $Q(w)$ are some polynomials. A general algorithm for solution of the integral equations for these polynomials of arbitrary orders was described in \cite{Alekseev:1988, Alekseev-Garcia:1996}. This algorithm leads to explicit form of solutions in a unified, but rather complicate form, and, as we shall see below, a large careful work is necessary for finding of appropriate choice of physical parameters which can simplify significantly the constructed solutions.

\subsection*{The linear singular integral equations}
For stationary axisymmetric fields outside their sources  the metric and electromagnetic vector potential can be considered in cylindrical coordinates in the form
\begin{equation}\label{LP-Metric}\begin{array}{l}
ds^2=g_{tt} dt^2+2 g_{t\varphi}dt d\varphi+g_{\varphi\varphi}d\varphi^2 -f(d\rho^2+dz^2)\\[1ex]
A_i=\{A_t,0,0,A_\varphi\},
\end{array}
\end{equation}
where  $x^i=\{t,\rho,z,\varphi\}$ and the metric components $g_{tt}$, $g_{t\varphi}$, $g_{\varphi\varphi}$ and $f$ as well as the components $A_t$ and $A_\varphi$ of the electromagnetic potential are functions of the coordinates $\rho$ and $z$ only. It is well known, that for the metric and electromagnetic fields (\ref{LP-Metric}) the symmetry reduced electrovacuum Einstein - Maxwell field equations decouple into two parts. One of these parts is a closed system of "dynamical" equations - the nonlinear partial differential equations for the functions $g_{tt}$, $g_{t\varphi}$, $g_{\varphi\varphi}$, $A_t$ and $A_\varphi$.
(We recall here that the Weyl cylindrical coordinates are defined so that $g_{tt} g_{\varphi\varphi}-g_{t\varphi}^2=-\rho^2$ and therefore, only two of these three metric components are unknown functions).  Another part is a pair of constraint equations which allow to determine in quadratures the conformal factor $f$ provided the solution for $g_{tt}$, $g_{t\varphi}$, $g_{\varphi\varphi}$, $A_t$ and $A_\varphi$ is already known.
Though an explicit calculation of these quadratures for the conformal factor $f$  represent usually a large technical difficulty, the principal problem is a construction of the solution of the "dynamical" equations with wanted physical and geometrical properties. That is why we concentrate below mainly on the construction of solution of the "dynamical" equations.

In accordance with a general scheme \cite{Alekseev:1985, Alekseev:1988},
the components of metric and electromagnetic potential (\ref{LP-Metric}), which satisfy the electrovacuum Einstein - Maxwell equations,
can be expressed in the form
\begin{eqnarray}\label{Rexpressions}
&&\begin{array}{l}g_{tt}=1-i
(R_t{}^\varphi-\overline{R}_t{}^\varphi)+\Phi_t \overline{\Phi}_t\\
\noalign{\medskip} g_{t\varphi}=-i (z-z_0)+i
(R_t{}^t+\overline{R}_\varphi{}^\varphi)+\Phi_t \overline{\Phi}_\varphi\\
\noalign{\medskip} g_{\varphi\varphi}=i
(R_\varphi{}^t-\overline{R}_\varphi{}^t)+\Phi_\varphi
\overline{\Phi}_\varphi
\end{array}
\quad
\begin{array}{l}
\hskip1ex{\cal E}=1-2 i R_t{}^\varphi\\[2ex]
\begin{pmatrix} \Phi_t\\ \Phi_\varphi\end{pmatrix} = 2 i
\begin{pmatrix} R_t{}^\ast\\ R_\varphi{}^\ast\end{pmatrix}
\end{array}
\end{eqnarray}
where $\{\Phi_t,\Phi_\varphi\}$ are the components of a complex  electromagnetic potential and  $\text{Re}\Phi_t=A_t$, $\text{Re}\Phi_\varphi = A_\varphi$; the functions $R_t^t$, $R_t^\varphi$, $R_t^\ast$, $R_\varphi^t$, $R_\varphi^\varphi$, $R_\varphi^\ast$ constitute a matrix which is determined by the integral  over the contour $L$ on the spectral plane
\begin{equation}\label{Rmatrix}
\mathbf{R}\equiv\begin{pmatrix}
R_t{}^t& R_t^\varphi & R_t^\ast\\
R_\varphi{}^t& R_\varphi^\varphi & R_\varphi^\ast
\end{pmatrix}\!=\!
\displaystyle\frac 1{i\pi}\!
\displaystyle\int\limits_L\!
[\lambda]_\zeta\!\!\left(\begin{array}{l}
1+i(\zeta-z_0)\mathbf{u}^\dagger(\zeta)\\[1ex]
-i(\zeta-z_0) \end{array}\!\!\right)
\otimes\bigl\{\bphi^{[\mathbf{1}]}(\zeta), \bphi^{[\mathbf{u}]}(\zeta),\bphi^{[\mathbf{v}]}(\zeta) \bigr\}d\zeta
\end{equation}
Here $\zeta\in L$ and the contour $L$ (unlike the general case) is a simple curve which starts from $w=\eta\equiv z-i\rho$, goes through the initial point $w=z_0$ and ends at the point $w=\xi\equiv z+i\rho$, where $\rho$ and $z$ are the well known cylindrical Weyl coordinates in which $g_{tt} g_{\varphi\varphi}-g_{t\varphi}=-\rho^2$, and $z_0$ determines the location of the point of normalization on the axis of symmetry $\rho=0$. In the above integral, $[\lambda]_{\zeta}$ denotes a jump (i.e. a half of the difference between left and right limits) at the point $\zeta\in L$ of a "standard" branching function \begin{equation}\label{lambda}
\lambda=\sqrt{(\zeta-z-i\rho)(\zeta-z+i\rho)/(\zeta-z_0)^2},\qquad \lambda(\rho,z,\zeta=\infty)=1.
\end{equation}
The functions $ \bphi^{[\mathbf{1}]}(\zeta)$, $ \bphi^{[\mathbf{u}]}(\zeta)$, $ \bphi^{[\mathbf{v}]}(\zeta)$ should satisfy to the decoupled linear singular integral equations with the same scalar kernels and different right hand sides \cite{Alekseev:1985,Alekseev:1988}:
\begin{equation}\label{LINEs}
-\dfrac 1{\pi i}
\vpint\limits_L\dfrac {[\lambda]_\zeta\,\mathcal{H}(\tau,\zeta)}{\zeta-\tau}\,
\begin{pmatrix}
\bphi^{[\mathbf{1}]}(\zeta)\\
\bphi^{[\mathbf{u}]}(\zeta)\\
\bphi^{[\mathbf{v}]}(\zeta)\end{pmatrix}\,
d\zeta=\begin{pmatrix}1\\\mathbf{u}(\tau)\\
\mathbf{v}(\tau)\end{pmatrix}
\end{equation}
where $\zeta,\tau\in L$; there is a Cauchy principal value integral in the left hand side, and the kernel function $\mathcal{H}(\tau,\zeta)$ in its integrand is
\[{\cal
H}(\tau,\zeta)=1+i(\zeta-z_0)
\bigl[{\bf u}^\dagger (\zeta)-{\bf u}(\tau)\bigr]+4(\zeta-z_0)^2 {\bf v}(\tau){\bf v}^\dagger (\zeta)\]
and everywhere below, we can put $z_0=0$ without any loss of generality. The coordinates $\rho$ and $z$ enter the equations (\ref{LINEs}) as the parameters which determine the location of the endpoints of the contour $L$ and as the arguments of the function $\lambda$, however, for simplicity we have not shown explicitly in (\ref{LINEs}) the dependence of the unknown functions $\bphi^{[\mathbf{1}]}$, $\bphi^{[\mathbf{u}]}$ and $\bphi^{[\mathbf{v}]}$ on these coordinates. We note also, that in general,  $\bphi^{[\mathbf{1}]}$, $\bphi^{[\mathbf{u}]}$ and $\bphi^{[\mathbf{v}]}$ as well as the right hand sides of (\ref{LINEs}) constitute the row-vectors, however, for a convenience we write (\ref{LINEs}) in a transposed form.

\section*{The Kerr - Newman field as a one-pole solution}
We begin our description of solution of the integral equation (\ref{LINEs}) with a simple case of a one-pole structure of the monodromy data functions:
\begin{equation}\label{One-pole}
\mathbf{u}(w)=\dfrac{u_0}{w-h},\qquad \mathbf{v}(w)=\dfrac{v_0}{w-h}.
\end{equation}
where $u_0$, $v_0$ and $h$ are arbitrary complex constants.
For these data the kernel function $\mathcal{H}(\tau,\zeta)/(\zeta-\tau)$ can be split into the singular and regular parts:
\begin{equation}\label{KernelSplit1}
\dfrac{\mathcal{H}(\tau,\zeta)}{\zeta-\tau}=\dfrac{1}{(\tau-h) (\zeta-\overline{h})}\left[\dfrac{P(\zeta)}{\zeta-\tau}+R(\zeta)\right]
\end{equation}
where the polynomials $P(\zeta)$ and $R(\zeta)$ possess the expressions ($z_0=0$):
\begin{equation}\label{PRpolynoms1}
\begin{array}{l}
P(\zeta)=\zeta^2[1-i (u_0-\overline{u_0})+4 v_0 \overline{v_0}]-\zeta [(1+i\overline{u_0})h+(1-i u_0)\overline{h}]+ h \overline{h}\\[1ex]
R(\zeta)=-(1+i \overline{u_0})\zeta+\overline{h}
\end{array}
\end{equation}
Assuming that $h\ne 0$, what means that the pole  is not located on the integration path $L$, we present the polynomial $P(w)$ in a factorized form
\begin{equation}\label{Pfactor1}
P(w)=P_0(w-w_1)(w-\widetilde{w}_1),\qquad
P_0=1-i(u_0-\overline{u_0})+4 v_0 \overline{v_0}
\end{equation}
The coefficients of $P(w)$ always are real and therefore, its roots $w_1$ and $\widetilde{w}_1$ are real or complex conjugated to each other. We parameterize these roots as
\begin{equation}\label{Roots1}
w_1=z_1+\sigma_1,\qquad \widetilde{w}_1=z_1-\sigma_1
\end{equation}
where $z_1$ is a real parameter while $\sigma_1$ can be real or pure imaginary. A comparison of (\ref{PRpolynoms1}) and (\ref{Pfactor1}) allows to express  $z_1$ and $\sigma_1$ in terms of $u_0$, $v_0$ and $h$, however, it is more convenient to use $z_1$ and $\sigma_1$ as new parameters and express some of the parameters $u_0$, $v_0$ and $h$ as functions of $z_1$,  $\sigma_1$ and others. We give the explicit expressions later, but now we concentrate on  solving of the integral equations (\ref{LINEs}).

As it can be concluded from the structure of the singular integral equations, their solutions for the monodromy data (\ref{One-pole}) should have the form
\begin{equation}\label{phis1}
\Bigl\{
\bphi^{[\mathbf{1}]}(w),\,\,
\bphi^{[\mathbf{u}]}(w),\,\,
\bphi^{[\mathbf{v}]}(w)\Bigr\}
=\dfrac{(w-\overline{h})}
{P(w)}\Bigl\{
w+X_0,\,\,
Y_0,\,\,
Z_0\Bigr\}
\end{equation}
where $X_0$, $Y_0$ and $Z_0$ are independent of the spectral parameter $w$, but they can depend on the coordinates $\rho$ and $z$. As we see from these expressions, the solutions of the integral equations (\ref{LINEs}) are rational functions of the spectral parameter and they have the poles coinciding with the roots of the polynomial $P(w)$.

To calculate the values of $X_0$, $Y_0$ and $Z_0$ explicitly, we substitute (\ref{One-pole}), (\ref{KernelSplit1}) and (\ref{phis1}) into the integral equations (\ref{LINEs}) and obtain the linear algebraic equations for $X_0$, $Y_0$ and $Z_0$ with rather complicate coefficients. These coefficients are linear combinations of the usual or  singular Cauchy principal value integrals of the form
\[\dfrac 1{\pi i}
\vpint\limits_L\dfrac {[\lambda]_\zeta\,\zeta^k}{\zeta-\tau}\,
\,
d\zeta,\qquad \dfrac 1{\pi i}
\int\limits_L\dfrac {[\lambda]_\zeta\,\zeta^k}{P(\zeta)}\,
\,
d\zeta
\]
where $\tau,\zeta\in L$ and $k\ge 0$ is some integer. Any integral of these types can be calculated explicitly using the elementary theory of residues. Indeed, the integrands in these integrals can be expressed as the jumps of analytical functions $\lambda(w)\,w^k/(w-\tau)$ and $\lambda(w)\,w^k/P(w)$ respectively. Therefore, these integrals can be expressed as the integrals over the closed curves $\mathcal{L}$ surrounding the contour $L$:
\begin{equation}\label{LIntegrals}
\dfrac 1{\pi i}
\vpint\limits_L\dfrac {[\lambda]_\zeta\,\zeta^k}{\zeta-\tau}\,
\,
d\zeta=\dfrac 1{2\pi i}
\int\limits_{\mathcal{L}}\dfrac {\lambda(\chi)\,\chi^k}{\chi-\tau}\,
\,
d\chi,
\qquad \dfrac 1{\pi i}
\int\limits_L\dfrac {[\lambda]_\zeta\,\zeta^k}{P(\zeta)}\,
\,
d\zeta=\dfrac 1{2\pi i}
\int\limits_{\mathcal{L}}\dfrac {\lambda(\chi)\,\chi^k}{P(\chi)}\,
\,
d\chi
\end{equation}
where $\chi\in\mathcal{L}$ and $\mathcal{L}$ surrounds $L$ in negative direction (i.e. so that the interior is to the right) and close enough to $L$ so that no poles of $P(w)$ are inside $\mathcal{L}$. The
function $\lambda(w)\,w^k/(w-\tau)$ is analytical  outside the contour $L$ and it can have only the pole at $w=\infty$, while the function $\lambda(w)\,w^k/P(w)$ is analytical outside  $L$, besides the poles which arise from zeros of $P(w)$ and possible poles at $w=\infty$. This allows us to transform the closed integration path $\mathcal{L}$ into a one approaching $w=\infty$ taking into account the inputs from the finite poles and therefore, in (\ref{LIntegrals}), each of the integrals  over $\mathcal{L}$ is equal to the sum of residues of its integrand at finite poles plus the residue at $w=\infty$.  Thus, for the integrals (\ref{LIntegrals}) we obtain
\begin{equation}\label{IntegralValues}
\begin{array}{l}
\dfrac 1{\pi i}
\vpint\limits_L\dfrac {[\lambda]_\zeta\,\zeta^k}{\zeta-\tau}\,
\,
d\zeta=-\displaystyle\sum\limits_{m=0}^k
(\lambda)_{k-m}\,\tau^m,\\[2ex]
\dfrac 1{\pi i}
\displaystyle\int\limits_L\dfrac {[\lambda]_\zeta\,\zeta^k}{P(\zeta)}\,
\,
d\zeta=\dfrac {\lambda (w_1)\,w_1^k}{P'(w_1)}+
\dfrac {\lambda (\widetilde{w}_1)\,\widetilde{w}_1^k}{P'(\widetilde{w}_1)}
-\left(\dfrac {\lambda (w)\,w^k}{P(w)}\right)_{-1}
\end{array}
\end{equation}
where $(\lambda)_{m-k}$ means the coefficient in front of $1/w^{k-m}$ in the inverse powers expansion of the function $\lambda(w)$ at $w=\infty$ and similarly, $(\ldots)_{-1}$ means the coefficient in front of $1/w$ in the inverse powers expansion of a function at $w=\infty$. The parameters $w_1$ and $\widetilde{w}_1$ are the roots of the polynomial $P(w)$, while $P'(w_1)$ and $P'(\widetilde{w}_1)$ mean a derivative of $P(w)$ with respect to $w$ at  $w=w_1$ or $w=\widetilde{w}_1$ respectively.

The substitution of (\ref{phis1}) and (\ref{KernelSplit1}) into (\ref{LINEs}) and the subsequent calculation of the contour integrals leads to the equations whose left and right hand sides, after multiplication by $(\tau-h)$, become polynomial functions of $\tau$. Equating the coefficients of these polynomials, we obtain the linear algebraic equations for $X_0$, $Y_0$ and $Z_0$ whose solution allows to obtain the explicit expressions for $\bphi$:
\begin{equation}\label{LINESolution1}
\Bigl\{
\bphi^{[\mathbf{1}]}(w),\,\,
\bphi^{[\mathbf{u}]}(w),\,\,
\bphi^{[\mathbf{v}]}(w)\Bigr\}
=\dfrac{(w-\overline{h})}
{P(w)}\Bigl\{
w+z-\dfrac{h+\Delta_1}{\Delta_0},\,\,
\dfrac{u_0}{\Delta_0},\,\,
\dfrac{v_0}{\Delta_0}
\Bigr\}
\end{equation}
where $\Delta_0$ and $\Delta_1$ possess the expressions
\begin{equation}\label{Deltas1}
\begin{array}{l}\Delta_0=-\dfrac{\lambda(w_1)} {P^\prime(w_1)}R(w_1)
-\dfrac{\lambda(\widetilde{w}_1)} {P^\prime(\widetilde{w}_1)}R(\widetilde{w}_1)
+1-\displaystyle\frac{1+i \overline{u_0}}{P_0}\\[2ex]
\Delta_1=-\dfrac{\lambda(w_1)} {P^\prime(w_1)}R(w_1)(w_1+z)
-\dfrac{\lambda(\widetilde{w}_1)} {P^\prime(\widetilde{w}_1)}R(\widetilde{w}_1)(\widetilde{w}_1+z)\\[1ex]
\phantom{\Delta_1=}
-\dfrac {1+i \overline{u_0}}{P_0^2}\left[\overline{h}(1-i u_0)+h(1+i \overline{u_0}) \right]+\dfrac{\overline{h}}{P_0}.
\end{array}
\end{equation}
and the polynomials $P(\zeta)$ and $R(\zeta)$ were defined in (\ref{PRpolynoms1}). Now we have to substitute these solutions of the integral equations into the integrands of quadratures (\ref{Rmatrix})
and calculate these quadratures using the same methods for calculation of these type of integrals  as it was described above. The components of metric and of electromagnetic potential, as well as the Ernst potentials for the constructing solution can be calculated then algebraically from the expressions (\ref{Rexpressions}).

The formal calculations for the monodromy data (\ref{One-pole}) described above lead us to the explicit but very complicate form of the solution. This solution can be simplified considerably using (i) a new and most appropriate set of parameters, (ii) some global gauge transformations -- $SL(2,R)$-rotations and rescalings of the Killing vectors $\xi_t$ and $\xi_\varphi$ with the corresponding linear transformations of $t$ and $\varphi$ coordinates and (iii) a set of more convenient coordinates (such as the prolate or oblate ellipsoidal coordinates) instead of the Weyl coordinates $\rho$ and $z$. In the remaining part of this section we describe these simplifications and show that this solution represents nothing more but the well known Kerr - Newman family.

At first, comparing the explicit expression (\ref{PRpolynoms1}) for $P(w)$ with its factorized form (\ref{Pfactor1}) and using (\ref{Roots1}), we obtain a set of relations between $z_1$,  $\sigma_1$ and the original set of independent parameters $u_0$, $v_0$ and $h$. Solving these relations we express $u_0$, $v_0$ and $h$ in terms of a new set of parameters which consists of the real parameters $m$, $a$, $b$, $z_1$ and a complex parameter $e$:
\begin{equation}\label{NewParameters1}
\begin{array}{lcccccl}
u_0=-i+\dfrac {i h}{w_1\widetilde{w}_1}(z_1+i a),&&&&&& h=z_1+m-i(a+b)
\\
v_0=\dfrac{i e}{2 h}\sqrt{P_0},&&&&&&
P_0=\dfrac{h \overline{h}}{w_1\widetilde{w}_1},
\end{array}
\end{equation}
and the parameter $\sigma_1$ is is related to the parameters $m$, $a$, $b$, $z_1$ and $e$ by the equation
\[\sigma_1^2=m^2+b^2-a^2- e \overline{e}\]
Now, for the choice of new coordinates we note, that besides a presence of $\rho$ and $z$ in the solution explicitly, its dependence of these coordinates also comes from the functions $\lambda(\rho,z,w)$ which should be calculated at the points $w=w_1$ and $w=\widetilde{w}_1$, i.e. at the roots of $P(w)$. Let us express these functions in the form
\[\lambda(\rho,z,w_1)=\dfrac{x_1-\sigma_1 y_1}{z_1+\sigma_1},\qquad
\lambda(\rho,z,\widetilde{w}_1)=\dfrac{x_1+\sigma_1 y_1}{z_1-\sigma_1}
\]
where $x_1$ and $y_1$ will be considered as new coordinates.
If we compare these expressions with the definition (\ref{lambda}), we obtain the relations between $\rho$, $z$ and $x_1$, $y_1$:
\[\rho=\sqrt{x_1^2-\sigma_1^2}\sqrt{1-y_1^2},\qquad z=z_1+x_1 y_1
\]
Finally, we have to make some gauge transformation of our solution:
$$
\begin{array}{lcclccl}
h_{tt}\to \displaystyle\frac{h_{tt}}{P_0}&&&{\cal E}\to
1+\displaystyle\frac{{\cal E}-{\cal E}_0+ 2\overline{\Phi}_0(\Phi-\Phi_0)}{P_0}&&&{\cal E}_0=1-2 i u_0\\
h_{t\varphi}\to h_{t\varphi} &&&\Phi\to\displaystyle\frac{\Phi-\Phi_0}{\sqrt{P_0}}&&&
\Phi_0=2 i v_0\\[2ex]
h_{\varphi\varphi}\to P_0 h_{\varphi\varphi}&&&
\widetilde{\Phi}\to\sqrt{P_0} \widetilde{\Phi}
\end{array}
$$
After this reparametrization, change of coordinates and gauge transformations we obtain the solution in a simple form. In particular, we obtain the Ernst potentials
\[\begin{array}{l}
\mathcal{E}=1-2\dfrac{(-i\epsilon_0 u_0+4\epsilon_0 v_0 v_0^\dagger)}{P_0\Delta_0}=1-\dfrac{2(m-i b)}{x_1+i a y_1+m-i b},\\[1ex] \Phi=-\dfrac{2 i v_0}{\sqrt{P_0}\Delta_0}=\dfrac{e}{x_1+i a y_1+m-i b}
\end{array}\]
which can be identified with those for the well known Kerr-Newman solution. The coordinates $x_1$ and $y_1$ are connected directly with  the polar spheroidal coordinates:
\[x_1=r_1-m,\qquad y_1=\cos \theta_1\]
and the real parameters $m$, $a$, $b$, $z_1$, real and imaginary part of $e$ can be identified respectively with the mass, angular momentum, NUT-parameter, electric and magnetic charges of the Kerr-Newman source (a black hole or a naked singularity).

\subsection*{The Reissner - Nordstr\"om  field as a static one-pole solution}
As it was already mentioned above, to obtain a static solution with the monodromy functions having only one pole, it is not necessary to make all of the calculations described above, but it would be enough to restrict from the beginning the monodromy data functions by the case, when these functions would take pure imaginary values on the real axis on the spectral plane. For this, it is necessary to choose the parameter $h$ to be real and the parameters $u_0$ and $v_0$ pure imaginary. In this case, as one can see from (\ref{NewParameters1}), we would have
\[h=\overline{h},\quad
u_0=-\overline{u_0},\quad
v_0=-\overline{v_0}
\qquad
\Longleftrightarrow \qquad
a=0,\quad
b=0,\quad
e=\overline{e}
\]
i.e. we obtain the Reissner - Nordstr\"om solution.

\section*{Superposing the fields of two Reissner - Nordstr\"om  sources}
In this section we describe the key points of the calculation of solutions for monodromy data having two simple poles on the spectral plane. This monodromy data can be represented as a superposition of two one-pole terms and therefore, in accordance with the previous section, it is naturally to expect that the corresponding solution  will describe the superposition of fields of two Reissner - Nordstr\"om  sources.
\subsection*{Structure of the monodromy data}
Thus, we begin with the choice of the monodromy data functions in the form
\begin{equation}\label{Two-pole}
\mathbf{u}(w)=\dfrac{u_1}{w-h_1}+\dfrac{u_2}{w-h_2},\qquad \mathbf{v}(w)=\dfrac{v_1}{w-h_1}+\dfrac{v_2}{w-h_2}.
\end{equation}
where $u_1$, $u_2$, $v_1$, $v_2$, $h_1$ and $h_2$ are arbitrary complex constants. However, to obtain a static solution, we specify this data  imposing the following constraints:
\begin{equation}\label{Staticity2}\begin{array}{lcclccl}
u_1=-\overline{u_1}&&& v_1=-\overline{v_1}&&&h_1= \overline{h_1}\\
u_2=-\overline{u_2}&&& v_2=-\overline{v_2}&&& h_2=\overline{h_2}\\
\end{array}
\end{equation}

\subsection*{Structure of the kernel and a new set of parameters}
In this case, the kernel  $\mathcal{H}(\tau,\zeta)/(\zeta-\tau)$ splits into the singular and regular parts:
\[
\dfrac{\mathcal{H}(\tau,\zeta)}{\zeta-\tau}=\dfrac{1}{(\tau-h) (\zeta-\overline{h})}\left[\dfrac{P(\zeta)}{\zeta-\tau}+R(\tau,\zeta)\right]
\]
where the polynomials $P(\zeta)$, $R(\tau,\zeta)$ are defined as (everywhere below $z_0=0$):
\begin{equation}\label{Ppolynom2}
\begin{array}{l}
P(\zeta)=h_1^2 h_2^2-2 h_1 h_2[(1-i u_1) h_2+(1-i u_2)h_1]\zeta+[(h_1+h_2)^2+2 h_1 h_2\\
\phantom{P(\zeta)=}
-4(h_1 v_2+h_2 v_1)^2-2 i (h_1^2 u_2+h_2^2 u_1)-4 i h_1 h_2(u_1+u_2)]\zeta^2\\
\phantom{P(\zeta)=}
+[2 h_1(-1+i u_1+2 i u_2+4 v_1 v_2+4 v_2^2)+
2 h_2(-1+2 i u_1+i u_2\\
\phantom{P(\zeta)=}+4 v_1 v_2+4 v_1^2)]\zeta^3
+\left[1-2 i (u_1+u_2)-4 (v_1+v_2)^2\right]\zeta^4\\[1ex]
R(\tau,\zeta)=R_0(\zeta)+R_1(\zeta) \tau\\[1ex]
R_0(\zeta)=
h_1 h_2(h_1 + h_2)\\
\phantom{R_0(\zeta)}-\left[h_1^2+3 h_1 h_2+h_2^2-i h_1(h_1+2 h_2) u_2-
i h_2(2 h_1+h_2) u_1\right]\zeta\\
\phantom{R_0(\zeta)}
+\left[2(h_1+h_2)(1-i u_1-i u_2-2 v_1 v_2)-i h_1 u_2-i h_2 u_1-4 h_1 v_2^2-4 h_2 v_1^2 \right] \zeta^2\\
\phantom{R_0(\zeta)}
+\left[-1+ 2 i (u_1+u_2)+4 (v_1+v_2)^2\right] \zeta^3\\
R_1(\zeta)=-h_1 h_2+(h_1+h_2-i h_1 u_2-i h_2 u_1)\zeta+(-1+i u_1+i u_2)\zeta^2
\end{array}
\end{equation}
Assuming that the pole of the monodromy data is not located on the integration path $L$ and therefore, $h_1 h_2\ne 0$, we present  $P(w)$ in a factorized form
\begin{equation}\label{Pfactor2}
P(w)=P_0(w-w_1)(w-\widetilde{w}_1) (w-w_2)(w-\widetilde{w}_2)
\end{equation}
The coefficients of the polynomial $P(w)$ always are real and therefore, its roots $w_1$, $\widetilde{w}_1$ and $w_2$, $\widetilde{w}_2$ are real or complex conjugated in pairs. We parameterize them as
\begin{equation}\label{Roots2}
\begin{array}{l}
w_1=z_1+\sigma_1\\
\widetilde{w}_1=z_1-\sigma_1
\end{array}\qquad
\begin{array}{l}
w_2=z_2+\sigma_2\\
\widetilde{w}_2=z_2-\sigma_2
\end{array}\qquad
\begin{array}{l}
h_1=z_1+\widetilde{m}_1\\
h_2=z_2+\widetilde{m}_2
\end{array} \qquad\qquad z_2=z_1+\ell
\end{equation}
where $z_1$ and $z_2$ (and therefore, $\ell$) are real parameters while $\sigma_1$ as well as $\sigma_2$ can be real or pure imaginary. A comparison of coefficients of different powers of the spectral parameter in (\ref{Ppolynom2}) and in (\ref{Pfactor2}) leads to a number of relations of the form
\begin{equation}\label{Vieta2}\begin{array}{r}
h_1^2 h_2^2=P_0 \mathcal{I}_4,\\[1ex]
2 h_1 h_2(h_1+h_2-i h_1 u_2-i h_2 u_1)=P_0\mathcal{I}_3,\\[1ex]
h_1^2+h_2^2+4 h_1 h_2-2 i(h_1+h_2)(h_1 u_2+h_2 u_1)-2 i h_1 h_2 (u_1+u_2)\\
-4 (h_1 v_2+h_2 v_1)^2=P_0 \mathcal{I}_2,\\[1ex]
2(h_1+h_2)-2 i (h_1 u_2+h_2 u_1)-2i(h_1+h_2)(u_1+u_2)\\
-8 (h_1 v_2+h_2 v_1) (v_1+v_2)=P_0 \mathcal{I}_1,\\[1ex]
1-2 i (u_1+u_2)-4 (v_1+v_2)^2=P_0.
\end{array}
\end{equation}
where we have introduced the notations
\[\begin{array}{l}
{\cal I}_1=w_1+\widetilde{w}_1+w_2+\widetilde{w}_2\\
{\cal I}_2=w_1 \widetilde{w}_1+w_1 w_2+w_1 \widetilde{w}_2+\widetilde{w}_1 w_2+\widetilde{w}_1 \widetilde{w}_2+w_2 \widetilde{w}_2\\
{\cal I}_3=\widetilde{w}_1 w_2 \widetilde{w}_2+
w_1 w_2 \widetilde{w}_2+
w_1 \widetilde{w}_1 \widetilde{w}_2+
w_1 \widetilde{w}_1 w_2\\
{\cal I}_4=w_1 \widetilde{w}_1 w_2 \widetilde{w}_2
\end{array}
\]
In terms of the parameters (\ref{Roots2}) these functions possess the expressions:
\[
\quad \begin{array}{lccl}
{\cal I}_1=2(z_1+z_2)&&&
{\cal I}_3=2 z_1 z_2(z_1+z_2)-2 z_2\sigma_1^2-2 z_1 \sigma_2^2\\
{\cal I}_2=z_1^2+4 z_1 z_2+z_2^2-\sigma_1^2-\sigma_2^2&&&
{\cal I}_4=(z_1^2-\sigma_1^2)(z_2^2-\sigma_2^2)
\end{array}
\]
The relations (\ref{Vieta2}) allow to express the roots of $P(w)$ (or the new parameters $z_1$, $z_2$ and $\sigma_1$, $\sigma_2$ in terms of $u_1$, $u_2$, $v_1$, $v_2$ and $h_1$, $h_2$, however, it is more simple and convenient to use $z_1$, $z_2$ and $\sigma_1$, $\sigma_2$ as new parameters and express $P_0$ and some of the parameters $u_1$, $u_2$, $v_1$, $v_2$ and $h_1$, $h_2$ as functions of $z_1$, $z_2$, $\sigma_1$, $\sigma_2$ and others. We present these expressions in the form
\begin{equation}\begin{array}{l}\label{Parametrization2}
u_1=\dfrac{2 i h_1 h_2^2 \widetilde{e}_1 \widetilde{e}_2 P_1 P_2+2 i h_1^3 \widetilde{e}_2^2 P_2^2 +
i h_1^2(h_1-h_2)(h_2^3{\cal I}_1-2 h_2^2 {\cal I}_2+3 h_2 {\cal I}_3-4 {\cal I}_4)}{2(h_1-h_2)^3 {\cal I}_4},\\[3ex]
u_2=\dfrac{-2 i h_1^2 h_2 \widetilde{e}_1 \widetilde{e}_2 P_1 P_2-2 i h_2^3 \widetilde{e}_1^2 P_1^2+
i h_2^2(h_1-h_2)(h_1^3{\cal I}_1-2 h_1^2 {\cal I}_2+3 h_1 {\cal I}_3-4 {\cal I}_4)}{2(h_1-h_2)^3 {\cal I}_4},\\[3ex]
v_1=-\dfrac{ i h_2 \widetilde{e}_1 P_1}{2(h_1-h_2)\sqrt{{\cal I}_4}},\qquad
v_2=-\dfrac{ i h_1 \widetilde{e}_2 P_2}{2(h_1-h_2)\sqrt{{\cal I}_4}}.
\end{array}
\end{equation}
where the parameters $\widetilde{e}_1$, $\widetilde{e}_2$ and $P_1$, $P_2$ are defined as follows. Solving (\ref{Vieta2}) leads to the expressions $P(h_1)$ and $P(h_2)$ which  must be non-negative:
\begin{equation}\begin{array}{l}\label{Ph1Ph2}
P(h_1)\equiv (h_1-w_1)(h_1-\widetilde{w}_1)(h_1-w_2)(h_1-\widetilde{w}_2)\ge 0\\
P(h_2)\equiv (h_2-w_1)(h_2-\widetilde{w}_1)(h_2-w_2)(h_2-\widetilde{w}_2)\ge 0
\end{array}
\end{equation}
Assuming for the case of real roots of $P(w)$ that they are numbered so that $\widetilde{w}_1<w_1<\widetilde{w}_2<w_2$, and that in any case, $z_1<z_2$, we conclude from (\ref{Ph1Ph2}) that $h_1$ and $h_2$ should be located outside the interval $(\widetilde{w}_1,w_1)$ if $w_1$ (and therefore, $\widetilde{w}_1$) is real and outside the interval $(\widetilde{w}_2,w_2)$ if $w_2$ (and therefore, $\widetilde{w}_2$) is real. This means that
\[(h_1-w_1)(h_1-\widetilde{w}_1)\equiv \widetilde{m}_1^2-\sigma_1^2\ge 0,\qquad (h_2-w_2)(h_2-\widetilde{w}_2)\equiv \widetilde{m}_2^2-\sigma_2^2\ge 0
\]
and therefore, there exist the real parameters $\widetilde{e}_1$ and $\widetilde{e}_2$ such that
\begin{equation}\label{e1te2t}
\widetilde{e}_1^2=\widetilde{m}_1^2-\sigma_1^2,\qquad \widetilde{e}_2^2=\widetilde{m}_2^2-\sigma_2^2
\end{equation}
and we allow to each of the parameters  $\widetilde{e}_1$ and $\widetilde{e}_2$ to be positive or negative in order to cover all opportunities to have in the expressions different signs in front of the square roots: $\pm\sqrt{P(h_1)}$ and $\pm\sqrt{P(h_2)}$. The parameters $P_1$ and $P_2$ in (\ref{Parametrization2}) are defined as positive parameters which have the following explicit expressions:
\begin{equation}\label{P1P2}
\begin{array}{l}
P_1=\sqrt{(h_1-w_2)(h_1-\widetilde{w}_2)}=
\sqrt{(\ell-\widetilde{m}_1)^2-\sigma_2^2}\\
P_2=\sqrt{(h_2-w_1)(h_2-\widetilde{w}_1)}=
\sqrt{(\ell+\widetilde{m}_2)^2-\sigma_1^2}
\end{array}
\end{equation}
The expressions (\ref{Parametrization2}) allow us to use instead  $u_0$, $u_1$, $v_0$, $v_1$ and $h_1$, $h_2$ the parameters $z_1$, $z_2$, $\widetilde{m}_1$, $\widetilde{m}_2$, $\widetilde{e}_1$, $\widetilde{e}_2$ (with $\ell=z_2-z_1>0$) which can take arbitrary real values, provided $P_1^2>0$ and $P_2^2>0$, in view of (\ref{P1P2}) and  (\ref{e1te2t}). Without loss of generality, we assume that if all roots are real, then $\sigma_1>0$, $\sigma_2>0$, and for the sources to be separated, we assume also that $\ell>\sigma_1+\sigma_2$.

\subsection*{Formal construction of the solution}
For solving of the integral equations (\ref{LINEs})
for the monodromy data (\ref{Two-pole}) - (\ref{Staticity2}), we substitute this data into the integral equation (\ref{LINEs}) and obtain the equations whose solution has the structure which is very similar
to that for the one-pole case (\ref{phis1}):
\begin{equation}\label{phis2}
\begin{pmatrix}
\bphi^{[\mathbf{1}]}(w)\\
\bphi^{[\mathbf{u}]}(w)\\
\bphi^{[\mathbf{v}]}(w)\end{pmatrix}
=\dfrac{(w-h_1)(w-h_2)}
{P(w)}\begin{pmatrix}
X_0+X_1 w+X_2 w^2\\
Y_0+Y_1 w+Y_2 w^2\\
Z_0+Z_1 w+Z_2 w^2\end{pmatrix}
\end{equation}
where the coefficients $X_k$, $Y_k$ and $Z_k$ ($k=0,1,2$) are independent of the spectral parameter $w$ and they are functions of coordinates and constant parameters. To find these coefficients explicitly, we have to substitute (\ref{phis2}) back into the integral equations, to calculate  the  corresponding integrals using the rules (\ref{IntegralValues}) and solve the linear algebraic equations for $X_k$, $Y_k$ and $Z_k$.
Using this solution, we find explicitly (using again the rules (\ref{IntegralValues})) all components (\ref{Rmatrix}) of the matrix $\mathbf{R}$ and calculate then pure algebraically the Ernst potentials and all metric and electromagnetic potential components in accordance with their general expressions (\ref{Rexpressions}).

\subsection*{Weyl cylindrical and bipolar coordinates}
During the calculation of the contour integrals (\ref{LIntegrals}), the coordinate dependence of the solution arises from calculation of residues of the integrands. It is easy to see, that the residue at infinity gives rise to the terms which are polynomial functions of Weyl coordinates $\rho$ and $z$, while the residues at finite poles --- the roots of $P(w)$ give rise to the terms which are proportional to the values
of the function $\lambda(\xi,\eta,w)$ at these roots: $\lambda_1=\lambda(\xi,\eta,w_1)$, $\widetilde{\lambda}_1=\lambda(\xi,\eta,\widetilde{w}_1)$, $\lambda_2=\lambda(\xi,\eta,w_2)$ and $\widetilde{\lambda}_2=\lambda(\xi,\eta,\widetilde{w}_2)$.
To have deal with these functions, we introduce bipolar coordinates $(x_1,y_1)$ and $(x_2,y_2)$ related to the corresponding pairs of spherical-like coordinates $(r_1,\theta_1)$ and $(r_2,\theta_2)$ centered on the symmetry axis $\rho=0$ at $z=z_1$ and $z=z_2$, so that $x_1=r_1-\widetilde{m}_1$, $y_1=\cos\theta_1$, $x_2=r_2-\widetilde{m}_2$, $y_2=\cos\theta_2$. The coordinates $(x_1,y_1)$ and $(x_2,y_2)$ are defined by the relations to $\lambda_1$, $\widetilde{\lambda}_1$, $\lambda_2$ and $\widetilde{\lambda}_2$
\[\begin{array}{lcccclccccl}
\lambda_1=\dfrac{x_1-\sigma_1 y_1}{z_1+\sigma_1}&&&&&
\widetilde{\lambda}_1=\dfrac{x_1+\sigma_1 y_1}{z_1-\sigma_1}&&&&&
\left\{\begin{array}{l}
\rho=\sqrt{x_1^2-\sigma_1^2}\sqrt{1-y_1^2}\\
z=z_1+x_1 y_1
\end{array}\right.
\\[3ex]
\lambda_2=\dfrac{x_2-\sigma_2 y_2}{z_2+\sigma_2}&&&&&
\widetilde{\lambda}_2=\dfrac{x_2+\sigma_2 y_2}{z_2-\sigma_2}&&&&&
\left\{\begin{array}{l}
\rho=\sqrt{x_2^2-\sigma_2^2}\sqrt{1-y_2^2}\\
z=z_2+x_2 y_2
\end{array}\right.
\end{array}
\]
while their relations to the Weyl coordinates given just above follow from the definition of the function $\lambda(\xi,\eta,w)$.

\subsection*{Physical parameters of the solution}
The calculations described above lead to rather complicate form of the solution expressed in terms of functions of bipolar coordinates $x_1$, $x_2$, $y_1$, $y_2$, the formal mass and charge parameters $\widetilde{m}_1$, $\widetilde{m}_2$, $\widetilde{e}_1$, $\widetilde{e}_2$ and the parameters $z_1$ and $z_2$ which characterize the location of the sources on the symmetry axis. However, it is more convenient to describe the structure of this solution using certain combinations of these parameters, which, as it will be shown later, represent physical parameters -- the masses of the sources $m_1$, $m_2$, their charges  $e_1$, $e_2$ calculated using the Gauss theorem and therefore, expressed in terms of the Komar-like integrals. These physical parameters are determined by the expressions
\[\begin{array}{lcccl}
m_1=\widetilde{m}_1+\dfrac{\widetilde{e}_1 \widetilde{e}_2 (\ell-\widetilde{m}_1+\widetilde{m}_2)}{P_1 P_2+\widetilde{e}_1 \widetilde{e}_2}&&&& e_1=\dfrac{\widetilde{e}_1 P_2(\ell-\widetilde{m}_1+\widetilde{m}_2)}{P_1 P_2+\widetilde{e}_1 \widetilde{e}_2}\\[2ex]
m_2=\widetilde{m}_2-\dfrac{\widetilde{e}_1 \widetilde{e}_2 (\ell-\widetilde{m}_1+\widetilde{m}_2)}{P_1 P_2+\widetilde{e}_1 \widetilde{e}_2}&&&&
e_2=\dfrac{\widetilde{e}_1 P_1+ \widetilde{e}_2 P_2}{ \ell-\widetilde{m}_1+\widetilde{m}_2}-e_1
\end{array}
\]
where  $P_1$, $P_2$ were  introduced in (\ref{P1P2}). The inverse relations are ($\gamma$ is defined below):
\[\begin{array}{l}
\widetilde{m}_1=\dfrac{(\ell+m_1+m_2)^2-K_0}{2(\ell+m_1+m_2)}\qquad
\widetilde{m}_2=\dfrac{(m_1+m_2)^2-\ell^2+K_0}{2(\ell+m_1+m_2)}\\[2ex]
\widetilde{e}_1^2=\dfrac12\left[(\ell+m_2)^2-m_1^2-K_0\right]
+ e_1(e_1+e_2)\\[0ex]
\widetilde{e}_2^2=\dfrac12\left[(\ell-m_1)^2-m_2^2-K_0\right]
+ (e_2+2\gamma)(e_1+e_2)+\dfrac{(m_1+m_2)K_0}{\ell+m_1+m_2}\\[1ex]
P_1^2=\dfrac12\left[(\ell-m_1)^2-m_2^2+K_0\right]
+ (e_2+2\gamma)(e_1+e_2)-\dfrac{(m_1+m_2)K_0}{\ell+m_1+m_2}\\[0ex]
P_2^2=\dfrac12\left[(\ell+m_2)^2-m_1^2+K_0\right]
+ e_1(e_1+e_2)\\[1ex]
\widetilde{e}_1 \widetilde{e}_2=-P_1 P_2 \left(\dfrac{(\ell+m_2)^2-m_1^2-K_0} {(\ell+m_2)^2-m_1^2+K_0}\right)\\[2ex]
K_0^2\equiv[(\ell+m_2)^2-m_1^2]^2+4 e_1(\ell+m_1+m_2)[e_2(\ell-m_1+m_2)+2 e_1 m_2]
\end{array}
\]
The parameters $\sigma_1$ and $\sigma_2$ also can be expressed in terms of $m_1$, $m_2$, $e_1$, $e_2$ and $\ell$:
\begin{equation}\label{sigmas2}\begin{array}{l}
\sigma_1^2=\widetilde{m}_1^2-\widetilde{e}_1^2=m_1^2-e_1^2+2 e_1\gamma,\\
\sigma_2^2=\widetilde{m}_2^2-\widetilde{e}_2^2=m_2^2-e_2^2-2 e_2\gamma,
\end{array}
\qquad
\text{where}\qquad \gamma=\dfrac{m_2 e_1-m_1 e_2}{\ell+m_1+m_2}
\end{equation}
Using these parameters, we obtain more symmetric and  simpler form of the solution. We introduce also, instead of $e_1$ and $e_2$,  two other parameters $q_1$ and $q_2$ such that:
\begin{equation}\label{q1q2}\begin{array}{l}
q_1=e_1-\gamma,\\
q_2=e_2+\gamma,
\end{array}\qquad \gamma=
\dfrac{m_2 q_1-m_1 q_2}{\ell},
\qquad \begin{array}{l}
\sigma_1^2=m_1^2+\gamma^2-q_1^2,\\[0.5ex]
\sigma_2^2=m_2^2+\gamma^2-q_2^2.
\end{array}
\end{equation}
Finally, after some long calculations and consideration of various possible forms of our solution, its rather short form was found.
In this form, all components of the solution are presented as functions of six parameters with only one constraint
\begin{equation}\label{Constraint2}
\left.\{m_1,\,m_2,\,q_1,\,q_2,\,\ell,\,\gamma\}
\qquad\vphantom{\displaystyle{\int}}\right\Vert\qquad
\gamma\ell=m_2 q_1-m_1 q_2
\end{equation}
which leaves only five  parameters to be independent.

\section*{5-parametric solution for interacting Reissner-Nordstr\"om sources}

For the monodromy data (\ref{Two-pole}),(\ref{Staticity2}), the corresponding solution is static and its metric and electromagnetic potential in cylindrical Weyl coordinates take the forms
\begin{equation}\begin{array}{l}\label{StaticMetric}
ds^2 = H dt^2-f(d\rho^2+dz^2)-\dfrac{\rho^2}{H} d\varphi^2, \\[1ex]
A_t = \Phi,\qquad A_\rho=A_z= A_\varphi=0,
\end{array}
\end{equation}
where $H$, $f$ and $\Phi$ are real functions of  $\rho$ and $z$. The calculations described in previous subsections
lead to the following structure of the functions $H$, $\Phi$ and $f$:
\begin{equation}\label{HFiDGF}
H=\dfrac{\mathcal{D}^2-\mathcal{G}^2 +\mathcal{F}^2} {(\mathcal{D}+\mathcal{G})^2},\qquad \Phi=\dfrac{\mathcal{F}} {\mathcal{D}+\mathcal{G}},\qquad f=\dfrac{f_0(\mathcal{D}+\mathcal{G})^2}
{(x_1^2-\sigma_1^2 y_1^2)
(x_2^2-\sigma_2^2 y_2^2)}
\end{equation}
where $\mathcal{D}$, $\mathcal{G}$, $\mathcal{F}$ are polynomial functions of bipolar coordinates with rather simple coefficients depending on the parameters of the solution:
\begin{equation}\label{DGF}
\begin{array}{l}
\mathcal{D}=x_1 x_2-\gamma^2 y_1 y_2\\
\phantom{\mathcal{D}=}+\delta\bigl[ x_1^2+x_2^2-\sigma_1^2 y_1^2-\sigma_2^2 y_2^2+2(m_1 m_2-q_1 q_2) y_1 y_2\bigr]\\[1ex]
\mathcal{G}=m_1 x_2+m_2 x_1+\gamma(q_1 y_1+q_2 y_2)\\
\phantom{\mathcal{G}=}+2\delta\left[m_1 x_1+m_2 x_2+y_1(q_2\gamma-m_1\ell)+y_2(q_1\gamma+m_2\ell)\right]\\[1ex]
\mathcal{F}=q_1 x_2+q_2 x_1+\gamma(m_1 y_1+m_2 y_2)\\
\phantom{\mathcal{F}=}+2\delta\bigl[q_1 x_1+q_2 x_2+y_1(m_2\gamma-q_1\ell)+y_2(m_1\gamma+q_2\ell)\bigr]
\end{array}
\end{equation}
In (\ref{HFiDGF}), $f_0$ is an arbitrary constant which should be chosen so that $f\to 1$ at spatial infinity and the parameter $\delta$ in (\ref{DGF})  is determined by the expression:
\begin{equation}\label{fdelta}
f_0=\dfrac 1{(1+2\delta)^2},\qquad
\delta=\dfrac{m_1 m_2-q_1 q_2}{\ell^2-m_1^2-m_2^2+q_1^2+q_2^2}
\end{equation}
In accordance with (\ref{q1q2}) and (\ref{Constraint2}), these equations give us an explicit expression of the solution in terms of five free real parameters $m_1$, $m_2$, $e_1$, $e_2$ and $\ell=z_2-z_1$. We consider this solution as depending on six parameters $m_1$, $m_2$, $q_1$, $q_2$, $\ell$ and $\gamma$ restricted by  the only one constraint (\ref{Constraint2}) and with the expressions (\ref{q1q2}) for $\sigma_1^2$, $\sigma_2^2$:
\begin{equation}\label{gammasigma}
\gamma\ell=m_2 q_1-m_1 q_2,\qquad
\sigma_1^2=m_1^2+\gamma^2-q_1^2,\qquad
\sigma_2^2=m_2^2+\gamma^2-q_2^2.
\end{equation}
This solution is asymptotically flat.
As it will be explained in the next section, the parameters $m_1$ and $m_2$  characterize the individual masses of the sources and the charges of these sources are $e_1=q_1+\gamma$ and $e_2=q_2-\gamma$ respectively, while the total mass  $m=m_1+m_2$ and the total charge $e=e_1+e_2$.

\section*{Physical parameters of the sources}
For electrovacuum space-time which admits a time-like Killing vector field $\xi=\partial_t$ we can write (following, for example, \cite{Alekseev:1988}) the "dynamical" part of the Einstein-Maxwell equations in the Kinnersley-like self-dual form
\begin{equation}\label{Self-dual}\nabla_k \overset +{\mathcal{H}}{}^{ik}=0,\qquad
\nabla_k \overset +{\mathcal{F}}{}^{ik}=0
\end{equation}
where the bivectors ${\mathcal{H}}{}_{ik}$ and ${\mathcal{F}}{}_{ik}$ are defined as follows ($\gamma=c=1$):
\begin{equation}\label{Bivectors}\begin{array}{lcclccl}
\overset +{\mathcal{H}}{}_{ik}\equiv \overset +{\mathcal{K}}{}_{ik}-2\overline{\Phi} {\overset + {\mathcal{F}}}{}_{kl},&&&
{\overset + {\mathcal{K}}}_{ik}=K_{ik}+\dfrac i2 \varepsilon_{iklm} K^{lm},&&&K{}_{ik}=\partial_i\xi_k-\partial_k\xi_i,\\
&&&{\overset + {\mathcal{F}}}_{ik}=F_{ik}+\dfrac i2 \varepsilon_{iklm} F^{lm},&&& F_{ik}=\partial_i A_k-\partial_k A_i
\end{array}
\end{equation}
and, due to (\ref{Self-dual}), these self-dual bivectors possess complex vector potentials, one of which determines the complex scalar function $\Phi$ which enters the expressions (\ref{Bivectors}):
\begin{equation}\label{HFPotentials}
{\overset + {\mathcal{H}}}{}_{ik}=\partial_i\mathcal{H}_k-\partial_k\mathcal{H}_i
\qquad
{\overset + {\mathcal{F}}}{}_{ik}=\partial_i\Phi_k-\partial_k\Phi_i
\qquad\Phi\equiv \xi^k\Phi_k
\end{equation}

The "dynamical" equations (\ref{Self-dual}) allow us to construct the "conserved" quantities -- the additive integral values which characterize the sources and which can be calculated as the integrals over the spherical-like 2-surfaces surrounding different parts of the sources on the space-like hypersurfaces slicing the space-time region outside the sources. For stationary axisymmetric spacetime outside the field sources we consider the space-like hypersurfaces $\Sigma_t:\,\, t=const$ where $t$ is the Killing parameter. Contraction the equations (\ref{Self-dual}) with the gradient $\partial_i t$, we obtain
\begin{equation}\label{Divergence}
\overset {(3)}{\nabla}_\delta (N_i\overset +{\mathcal{H}}{}^{i \delta})=0,\qquad
\overset {(3)}{\nabla}_\delta (N_i\overset +{\mathcal{F}}{}^{i \delta})=0
\end{equation}
where $\delta=1,2,3$; $\overset {(3)}{\nabla}$ is a covariant derivative with respect to a three-dimensional metric on the hypersurfaces $\Sigma_t$ and $N_i$ is a unit time-like future-directed normal to these hypersurfaces. Integrating the above equations on $\Sigma_t$ over a three-dimensional  region between a sphere of a large radius $B_\infty$ located in the asymptotically Minkowski region and by a closed surface $B_s$ surrounding the sources, and applying the Gauss theorem, we obtain
\begin{equation}\label{BIntegrals}
\displaystyle\int\limits_{B_s} ({\overset +{\mathcal{H}}} {}_{i k} N^i n^k) d^2\sigma=\displaystyle\int\limits_{B_\infty} ({\overset +{\mathcal{H}}} {}_{i k} N^i n^k) d^2\sigma,\qquad
\displaystyle\int\limits_{B_s} ({\overset +{\mathcal{F}}} {}_{i k} N^i n^k) d^2\sigma=\displaystyle\int\limits_{B_\infty} ({\overset +{\mathcal{F}}}{}_{i k} N^i n^k) d^2\sigma
\end{equation}
where $n_k$ is a unit vector tangent to the hypersurface $\Sigma_t$ and representing the outward normal to the corresponding boundary $B_s$ or $B_\infty$ and $d^2\sigma$ is the area element on these boundaries.
At first, we calculate the integrals over $B_\infty$.

For any stationary axisymmetric  asymptotically flat electrovacuum solution of Einstein - Maxwell equations, the metric and electromagnetic potential of the form (\ref{LP-Metric}) at spatial infinity admit the expansions (the NUT parameter, the total magnetic charge and the additive constants in $A_t$ and $A_\varphi$ are assumed to vanish):
\[\begin{array}{lccl}
g_{tt}=1-\dfrac{2 m}r+O(\dfrac1{r^2})&&&f=1+\dfrac{2 m}r+O(\dfrac1{r^2})\\
g_{t\varphi}=\dfrac{2 a m}{r} \sin^2\theta+O(\dfrac1{r^2})&&& A_t=\dfrac{e}r+O(\dfrac1{r})\\
g_{\varphi\varphi}=-r^2 \sin^2\theta-2 m r \sin^2\theta+O(r^0)&&&A_\varphi=O(\dfrac1{r})
\end{array}
\]
where $\rho=r\sin\theta$, $z=z_\ast+r\cos\theta$ and the constants $m$, $a$ and $e$ mean the total mass, total angular momentum per unit mass and the  total electric charge respectively. For the complex vector potentials $\mathcal{H}_i$ and $\Phi_i$ introduced in (\ref{HFPotentials}) for $r\to\infty$ in the coordinates $\{t,\rho,z,\varphi\}$ we obtain the expansions
\begin{equation}\label{SDpotentials}
\begin{array}{l}
\mathcal{H}_i=\{-\dfrac{2 m}{r}+O(\dfrac1{r^2}),\,\,0,\,\,0,\,\,-2 i m\cos\theta+O(\dfrac1{r})\}\\[1ex]
\Phi_i=\{\dfrac{e}{r}+O(\dfrac1{r^2}),\,\,0,\,\,0,\,\, i e \cos\theta+O(\dfrac1{r})\}
\end{array}
\end{equation}
The components of the vector $N^i$ in these coordinates take the form
\[N^i=\{1+\dfrac{m}{r}+O(\dfrac1{r^2}),\,\,0,\,\,0,\,\, \dfrac{2 a m}{r^3}+O(\dfrac1{r^4})\}
\]
For a sphere of a large radius, the spatial unit normal vector in the leading order is $n^i\partial_i=\partial/\partial r$ and in the limit $r\to\infty$ for the integrals (\ref{BIntegrals})
over $B_\infty$ we obtain
\[\displaystyle\int\limits_{B_\infty} ({\overset +{\mathcal{H}}} {}_{i k} N^i n^k) d^2\sigma=-8\pi m,\qquad
\displaystyle\int\limits_{B_\infty} ({\overset +{\mathcal{F}}}{}_{i k} N^i n^k) d^2\sigma=4\pi e
\]
where $m$ and $e$ are the total mass and charge of the field configuration. This allows us to express (in accordance with (\ref{BIntegrals})) the total mass and charge of a system of sources in terms of the integrals over the surface $B_s$ surrounding the sources:
\begin{equation}\label{MtQt}
m=-\dfrac 1{8\pi}\displaystyle\int\limits_{B_s} ({\overset +{\mathcal{H}}} {}_{i k} N^i n^k) d^2\sigma,\qquad
e=\dfrac 1{4\pi}\displaystyle\int\limits_{B_s} ({\overset +{\mathcal{F}}}{}_{i k} N^i n^k) d^2\sigma
\end{equation}
Now we transform the integrands in (\ref{MtQt}) using the self-duality of ${\overset +{\mathcal{H}}}$ and ${\overset +{\mathcal{F}}}$:
\begin{equation}\label{klvectors}
\left.\begin{array}{l}
\displaystyle\int\limits_{B_s}{\overset{\hskip0.5ex +}{\mathcal H}}{}^{ij}N_i n_j d^2\sigma
=i\displaystyle\int\limits_{B_s}{\overset{\hskip0.5ex +}{\mathcal H}}{}_{ij} k^i l^j d^2\sigma\\[3ex]
\displaystyle\int\limits_{B_s}{\overset{\hskip0.5ex +}{\mathcal F}}{}^{ij}N_i n_j d^2\sigma
=i\displaystyle\int\limits_{B_s}{\overset{\hskip0.5ex +}{\mathcal F}}{}_{ij} k^i l^j d^2\sigma
\end{array}\qquad\right\Vert\qquad\begin{array}{l}
\varepsilon^{ijkl} N_k n_l=k^i l^j-k^j l^i\\[2ex]
\varepsilon^{ijkl} N_k n_l k_i l_j=1
\end{array}
\end{equation}
Here we introduced two spatial unit vectors $k^i$ and $l^i$ which are orthogonal to each other and tangent to 2-surface $B_s$. These vectors determine the orientation of the 2-surface $B_s$ so that the spatial basis $\{n^i,\,k^j,\,l^k\}$ is positively oriented. Choosing the 2-surface $B_s$ to be axially symmetric, we specify
the choice of the vectors $k^i$ and $l^j$ in the tangent space of $B_s$ so that $l^j$ would be the rotational Killing vector $\partial/\partial_\varphi$ and  $k^i$ is tangent to a curve $\mathcal{L}$ on $B_s$ with $\varphi=const$:
\[k^i=\{0,\,\dfrac{d\rho}{d\ell},\,\dfrac{dz}{d\ell},\, 0\}\qquad\qquad l^i=\dfrac{1}{\sqrt{-g_{\varphi\varphi}}} \{0,\,0,\,0,\,1\}
\]
where the parameter $\ell$ is the length on the curve $\mathcal{L}$ and the direction on  $\mathcal{L}$ is chosen so that it goes from some point on the positive part of the $z$-axis to some point on its negative part (provided the point $\rho=z=0$ is located somewhere inside $B_s$). The element of the area on $B_s$ is $\sqrt{-g_{\varphi\varphi}} d\ell d\varphi$. After the integration over $\varphi$ we observe that the integrands of the remaining contour integrals over $\mathcal{L}$ are the differentials of the $\varphi$-components of the potentials of self-dual bivectors ${\overset +{\mathcal{H}}}$ and ${\overset +{\mathcal{F}}}$:
\begin{equation}\label{MtotalQtotal}
\begin{array}{l}
m=-\dfrac {i}{4}\displaystyle\int\limits_{\mathcal{L}}(\partial_\mu {\mathcal H}{}_\varphi)d x^\mu=\dfrac {i}{4}\left[({\mathcal H}{}_\varphi)_+-({\mathcal H}{}_\varphi)_-\right]\\[3ex]
e=\dfrac {i}{2}\displaystyle\int\limits_{\mathcal{B_s}L}(\partial_\mu \Phi{}_\varphi)d x^\mu=-\dfrac {i}{2}\left[(\Phi_\varphi)_+-(\Phi_\varphi)_-\right]
\end{array}
\end{equation}
where $(\ldots)_+$ means the value of a potential at the beginning of  $\mathcal{L}$, i.e. on the positive part of the $z$-axis, and $(\ldots)_-$ -- its value on the negative part of this axis.

It is important to note that $\mathcal{H}_\varphi$ and   $\Phi_\varphi$  are constant along the regular parts of the symmetry axis outside the sources and therefore, the integrals (\ref{MtotalQtotal}) do not change if we deform the surface of integration $B_s$ surrounding the sources
in the domain of regularity of the solution.

Now we consider the case, in which the source of the field consists of two parts (black holes, naked singularities or extended bodies) separated by some segment of the symmetry axis. In this case, we can deform the surface $B_s$ surrounding the sources into two spheres each surrounding one of these parts of the sources and a thin tube surrounding the intermediate part of the axis between the sources. If  $\mathcal{H}_\varphi$ and   $\Phi_\varphi$ are constant on the intermediate part of the axis, the integral over the thin tube vanish and we obtain that the integrals for the total mass and the total charge are expressed as the sums of the integrals of the same type over the spheres surrounding each of the sources. In particular, for the sources consisting of two parts, as it is in our solution for two Reissner - Nordstr\"om sources, we obtain
\[m=m_1+m_2,\qquad e=e_1+e_2\]
where $m_1$, $m_2$ and  $e_1$, $e_2$ can be interpreted respectively as the masses (energies) and charges of the corresponding parts of the source. To conclude this section, we mention that rather tedious calculations show that the parameters $m_1$, $m_2$ and  $e_1$, $e_2$ in our electrostatic solution presented in the previous section and describing the field of two Reissner - Nordstr\"om sources, coincide with the values of the integrals (\ref{MtotalQtotal}) calculated for each part of the sources and therefore, these parameters can be interpreted as the additive masses and charges of these interacting sources.

\section*{Equilibrium of two Reissner - Nordstr\"om sources}
The 5-parametric solution (\ref{StaticMetric}) - (\ref{gammasigma}) presented above is asymptotically flat and the space-time geometry is regular far enough from the sources. However, for an arbitrary choice of parameters $m_1$, $m_2$, $q_1$, $q_2$ and $\ell$, this solution may have some physically not reasonable singularities ("struts") on the part of the axis of symmetry between the sources. Like at the vertex of the cone, at these points, the local Eucledean properties of space-time may be violated so that  on the space-time sections $t=const$, $z=const$  the ratio of the length $\mathcal{L}$ of a small circle, surrounding the axis of symmetry $\rho=0$ and contracting to the point $z$ of this axis, to its radius $\mathcal{R}$ multiplied by $2\pi$ is not equal to a unit.
The constraint imposed on the parameters which provide the absence of such conical singularities plays the role of equilibrium condition because it allows to select a physically acceptable solution in which these singularities are absent and the sources are in the equilibrium because of the balance between their gravitational and electromagnetic interactions.

As it follows from elementary considerations, the limiting value of the square of the ratio $(2\pi\mathcal{R})/\mathcal{L}$ for $\mathcal{R}\to 0$ is equal to the value of the product $f H$ at the point $z$ of the axis. However, the field equations imply that on the regular parts of the axis of symmetry the product $ f(\rho=0,z) H(\rho=0,z)$ does not depend on $z$ and therefore, it is a constant. However, the values of this constant can be different on different parts of the axis of symmetry separated by the sources, and the condition $f H=1$ of the absence of conical singularities should be satisfied at each of these parts of the axis of symmetry.

In our solution with two separated sources, there exists three parts of the axis of symmetry where the condition of the absence of conical points should be considered. There are two semi-infinite parts of the axis  outside the sources. On the negative part $L_-$, we have $-\infty< z <z_1-\sigma_1$ (if the first source is a black hole) or $-\infty< z <z_1$ (if the first source is a naked singularity) and on this part
$x_1=z_1-z$, $x_2=z_2-z$, $y_1=-1$, $y_2=-1$. On the positive part $L_+$ of the axis, $z_2+\sigma_2< z< \infty$ (if the second source is a black hole) or $z_2< z< \infty$  (if the second source is a naked singularity) and we have there $x_1=z-z_1$, $x_2=z-z_2$, $y_1=1$, $y_2=1$. It is easy to see that on these parts of the axis the condition $f H=1$ is satisfied for any choice of free parameters of our solution. However, it is not the case on the part $L_0$ of the axis between the sources. On $L_0$ we have $z_1+Re(\sigma_1)< z< z_2-Re(\sigma_2)$ and $x_1=z-z_1$, $x_2=z_1+\ell-z$, $y_1=1$, $y_2=-1$ and the product $f H$ also takes there some constant value whose equality to a unit give us the equilibrium condition:
\begin{equation}\label{Equilibrium}
m_1 m_2-q_1 q_2=0
\end{equation}
It is interesting to note that this equilibrium condition looks just like the Newtonian condition of equilibrium of two charged point-like masses, but in the case of General Relativity this condition relates the masses of the Reissner-Nordstr\"om sources not with their charges $e_1$ and $e_2$, but with the parameters $q_1$ and $q_2$ whose expressions in terms of masses $m_1$, $m_2$ and charges $e_1$, $e_2$ depend also on the $z$-distance $\ell$ separating the sources.

The equilibrium condition (\ref{Equilibrium}) allows  also to simplify the solution (\ref{StaticMetric}) - (\ref{gammasigma}) and leads to the 4-parametric family which describes the superposed field of two Reissner - Nordstr\"om sources in equilibrium. This solution has been presented in our previous short paper \cite{Alekseev-Belinski:2007}. Here we persent it in a bit different form using $m_1$, $m_2$, $q_1$, $q_2$, $\gamma$ as the basic set of parameters, such that the first four of them should satisfy the equilibrium condition (\ref{Equilibrium}). For this solution the metric functions have the same expressions (\ref{HFiDGF}) where the polynomials $\mathcal{D}$, $\mathcal{G}$ and $\mathcal{F}$
have the expressions
\[
\left.\begin{array}{l}
\mathcal{D}=x_1 x_2-\gamma^2 y_1 y_2 \\[1ex]
\mathcal{G}=m_1 x_2+m_2 x_1+\gamma(q_1 y_1+q_2 y_2) \\[1ex]
\mathcal{F}=q_1 x_2+q_2 x_1+\gamma(m_1 y_1+m_2 y_2)
\end{array}\quad\right\Vert\hskip1ex
\begin{array}{l}
{}\\[1ex]
m_1 m_2=q_1 q_2\\[1ex]
{}\\[1ex]
\end{array}
\]
and $f_0=1$. The other physical parameters -- the $z$-distance $\ell$ separating the sources and their charges are determined by the expressions
\[\ell\equiv z_2-z_1=(m_2 q_1-m_1 q_2)/\gamma,\qquad e_1=q_1+\gamma,\quad e_2=q_2-\gamma
\]
The independent parameters $m_1$, $m_2$, $q_1$, $q_2$, $\gamma$ of this solution should be chosen so that the sources would be separated actually by some positive distance, i.e. $\ell> Re(\sigma_1)+Re(\sigma_2)$.

This concludes our description in this paper of some fragments of the monodromy transform approach and of the procedure for solution of the corresponding singular integral equations for stationary axisymmetric electrovacuum fields with simple rational monodromy data, which leads to the construction of the  5-parametric family of solutions for the field of two interacting Reissner-Nordstr\"om sources, and of the derivation from this solution of the 4-parametric family of solutions for the fields of equilibrium configurations of these sources. Some properties of these equilibrium field configurations have been discussed in \cite{Alekseev-Belinski:2007}, however, a more detail analysis of physical and geometrical properties of these configurations, such as the structure of the superposed fields, influence on the geometry of horizons and on the space-time geometry inside the horizon of the external gravitational and electromagnetic fields created by another source, tidal influence of these fields on the structure of naked singularities, stability of equilibrium and probably, some others, are expected to be the subject of our next publications.

\section*{Acknowledgements}
The authors are thankful to the Organizing Committee of
MG11 for the invitation to submit this paper devoted to mathematical aspects of solving of Einstein-Maxwell equations to the Proceedings of this Meeting.

GAA is thankful to ICRAnet for the financial support and hospitality during his visit to ICRAnet (Pescara, Italy) during June 2007, when this paper was started. The work of GAA was also supported in parts by the Russian Foundation for Basic Research (grants 05-01-00219, 05-01-00498, 06-01-92057-CE) and the programs "Mathematical Methods of Nonlinear Dynamics" of the Russian Academy of Sciences, and "Leading Scientific Schools" of Russian Federation (grant NSh-4710.2006.1).

\end{document}